\documentstyle[12pt,aasms4]{article}
\begin{document}

\def\kms{\rm km~s$^{-1}$}
\def\erg{\rm erg}
\def\ergs{\rm ergs s$^{-1}$}
\def\Lx{L_{X}}
\def\Lb{L_{B}}
\def\fb{f_{B}}
\def\fx{f_{X}}
\def\Mbh{M_{\rm BH}}
\def \eg           {{e.g.}}
\def \date         {\ifcase\month \message{zero} \or
                    January \or February \or March \or April \or May \or June 
                    \or July \or 
                    August \or September \or October \or November \or 
                    December \fi
                    \space\number\day, \number\year}

\def\Halpha{H$\alpha$}
\def\Hbeta{H$\beta$}
\def\OI{[OI]$\lambda$6300}
\def\OII{[OII]$\lambda$3727}
\def\OIIIone{[OIII]$\lambda$4959}
\def\OIIItwo{[OIII]$\lambda$5007}
\def\OIII{[OIII]$\lambda$4959,5007}
\def\NII{[NII]$\lambda$6584}
\def\SIIone{[SII]$\lambda$6716}
\def\SIItwo{[SII]$\lambda$6731}
\def\SIIthree{[SII]$\lambda$6716+6731}
\def\SII{[SII]$\lambda$6716,6731}

\def\remark#1{{\bf (#1)}}

\title{An X-ray Luminous, Dwarf Seyfert Companion of Mrk~273}
\author{X.-Y. Xia$^{1,2}$,Th. Boller$^{3}$, H. Wu$^{4}$, Z.-G. Deng$^{5,2}$,
Y. Gao$^{6}$, Z.-L. Zou$^4$, S. Mao$^{2}$, G. B\"orner$^{2}$}
\altaffiltext{1}{
	Dept. of Physics, Tianjin Normal University, 300074 Tianjin, PRC}
\altaffiltext{2}{Max-Planck-Institut f\"ur Astrophysik,
	Karl-Schwarzschild-Str. 1, 85740 Garching, Germany}
\altaffiltext{3}{Max-Planck-Institut f\"ur Extraterrestrische Physik, 
	Karl-Schwarzschild-Str. 1, 85740 Garching, Germany}
\altaffiltext{4}{Beijing Astronomical Observatory, 
	Chinese Academy of Sciences, 100080 Beijing, PRC}
\altaffiltext{5}{Dept. of Physics, Graduate School, 
	Chinese Academy of Sciences, 100039 Beijing, PRC}
\altaffiltext{6}{Dept. of Astronomy, University of Illinois, 1002
	W. Green St., Urbana, IL 61801, USA}

\received{ }
\accepted{ }

\begin{abstract}

We report the discovery of the brightest X-ray source
hosted by a faint ($M_B=-16$)
dwarf galaxy in the immediate vicinity of
the ultraluminous IRAS merging galaxy Mrk 273.
The dwarf galaxy, $1.3\arcmin$ away from Mrk 273, is 
at the tip of a faint northeast plume of Mrk 273. 
Its spectrum exhibits strong [OIII], \Halpha, [NII] emission lines,
which establish the redshift of the dwarf galaxy,
$z=0.0376$, the same as that of Mrk 273.
The emission line ratios are typical of Seyfert galaxies.
The X-ray emission is consistent with 
a point-like source coincident with the center of the dwarf galaxy.
The intrinsic X-ray luminosity, $6.3\times 10^{41}$ \ergs, in the 0.1--2.4 keV
energy range, is about seven times larger than the B band luminosity.
The X-ray spectrum of the source can be fit with a power-law.
All the evidence is consistent with the source being a Seyfert galaxy.
It is mysterious why out of $\sim 10$ faint objects
in the same field only one is detected by ROSAT and its
ratio of soft X-ray to optical luminosity is as high as those for
BL Lac objects and few active galactic nuclei (AGNs).
If there is a population of such dwarf AGNs
hidden as companions of major merger galaxies (such as Mrk 273), they
may contribute to the luminosity function of AGNs and the cosmic X-ray
background at the faint end.
\end{abstract}


\clearpage

\section{Introduction}

During the cross-correlation of ultraluminous IRAS galaxies (ULIRGs) selected
from 1.2Jy (Fisher et al. 1995) and QDOT all sky redshift survey 
(Lawrence et al. 1998)
with X-ray sources from ROSAT All Sky Survey and archive data, we found that
some ULIRGs have soft X-ray companions within a few
arcminutes of each source. The optical counterparts of these companions
are much fainter than the primary target objects (Xia et al. 1998). 
This phenomenon was first mentioned by 
Turner, Urry \& Mushotzky (1993) using ROSAT PSPC observations of
six Seyfert 2 galaxies. However it was unclear
whether the soft X-ray companions are physically 
associated with the target galaxies or just due to chance alignment
(see below). The origin of
the soft X-ray emission from these companions is not well understood:
presumably it is closely related to the interacting
and/or merging process that is occurring in almost all ULIRGs
(\eg, Sanders \& Mirabel 1996; Wu et al. 1998 and references therein).

To understand the physical connection between the soft X-ray companions with
the primary targets, we initiated an optical spectroscopic program to observe
the faint counterparts of soft X-ray
companions in Arp 220, Mrk 231 and Mrk 273. For Arp 220 ($z=0.018$), 
the counterpart of the soft X-ray companion is likely a background group
or cluster with redshift of about 0.1. For Mrk 231 ($z=0.042$), 
the counterparts are very likely background
QSOs with redshift larger than 1. However, for Mrk 273 ($z=0.0378$),
our observations show that its soft X-ray companion is physically 
associated with Mrk 273 because the faint counterpart at the center
of soft X-ray emission has the same redshift as Mrk 273. The
counterpart, which we term as Mrk 273x,
turns out to be a nucleated dwarf galaxy. 
The few dwarf galaxies detected previously in X-ray have
X-ray luminosities $\lesssim 10^{39}$~\ergs~ (e.g., 
Gizis, Mould \& Djorgovski 1993 for Fornax; Papaderos et al. 1994 for BCD VII ZW 403;
Martin \& Kennicutt 1995 for NGC 5253; Hensler et al. 1996 for NGC 1705).
The search for low-luminosity Seyfert nuclei in nearby galaxies
by Ho et al. (1996) did not find any Seyfert nucleus for sources 
fainter than $M_B \approx -17.5$.
The optical counterparts for broad line AGNs
selected from Einstein Observatory Extended Medium
Sensitivity Survey are generally brighter than $M_B=-18$ (Della Ceca
et al. 1996). In contrast,
the soft X-ray luminosity of Mrk 273x is $6.3 \times 10^{41}$ \ergs,
more than two orders of magnitude higher than those from other dwarf 
galaxies, and is by far the brightest observed to date. In \S 2 and 3,
we report the optical and X-ray observations of Mrk 273x.
Combined with the recent VLA 21cm continuum
and HST WFPC2 archive observations, we demonstrate that this object is a
unique dwarf AGN. We discuss the implications of Mrk 273x in \S 4.
Throughout this paper, we use a Hubble constant
$H_0=50~\rm km\ s^{-1}\ Mpc^{-1}$.

\section{Observations and Data Reduction}

The observations with the Position Sensitive Proportional Counter (PSPC) and
High Resolution Imager (HRI) instruments on board the ROSAT X-ray satellite
for the Mrk 273 field were carried out during May and June 1992
as archived. The total exposure times are 37.6ksec and 19.3ksec for the PSPC and HRI
respectively. Fig. 1 (plate) shows the soft X-ray contours from the ROSAT HRI
image overlaid on the optical image scanned 
from an R film of the second Palomar Observatory Sky Survey (POSS II).
One can see a strong soft X-ray companion
$1.3\arcmin$ (84 kpc in projected distance)
to the northeast of Mrk 273. At the center of the soft X-ray companion,
a faint optical counterpart is clearly visible.
This counterpart (which we call Mrk 273x),
is at the tip of the northeast plume of Mrk 273.
The apparent B and R magnitudes for Mrk 273x 
are about 20.8 and 19.6, respectively,
from the USNO-A1.0 catalog (Monet 1996).
In order to know if the soft X-ray companion is physically associated with
Mrk 273, we observed this object on 12 April, 1997 using a Zeiss
universal spectrograph mounted on the 2.16m telescope at Xinglong Station
of Beijing Astronomical Observatory (BAO). A Tektronix 1024$\times$1024 CCD was
used giving a wavelength coverage
of 3500\AA\ to 8100\AA\ with a grating of 195\AA/mm and
a spectral resolution of 9.3\AA\ (2 pixels). Wavelength calibration was
carried out using an Fe-He-Ar lamp; the resulting wavelength accuracy is 
better than 1\AA. KPNO standard stars were observed to perform flux 
calibrations.

The optical spectral data reduction was performed using IRAF packages at
BAO. The CCD data reduction includes standard procedures such as
bias subtraction, flat fielding and cosmic ray removal.
All the X-ray data reduction was carried out at the
Max-Planck-Institut f\"ur Extraterrestrische Physik using the EXSAS software.
The detected photon counts from the companion of Mrk 273
by PSPC and HRI are 444 and 61, respectively. The HRI data can be used
to determine the image position with 
an accuracy $\leq 5\arcsec$. The PSPC data, on the other hand,
allow one to do a spectral analysis in the 0.1--2.4 keV 
energy band, obtain the radially averaged intensity
profiles (deconvolved with
the PSPC point-spread function) and test time variability. 

\section{Results and Analysis}

\subsection{The Optical Spectroscopy of Mrk~273x}

The optical spectrum for Mrk 273x is shown in Fig. 2. 
The measurement of line emissions was performed using
IRAF tasks ``splot'' and ``ngaussfit''. 
Since \Halpha\ emission is heavily blended with \NII, a 
double-Gaussian fit was employed. The inset in Fig. 2 shows
that the fit is quite satisfactory.
The redshift of Mrk 273x determined from the 
\Halpha\  emission line is
$z=0.0376$, nearly identical to that of Mrk 273. From the apparent
magnitudes, $B=20.8$ and $R=19.6$, we obtain
absolute magnitudes $-16$ and $-17.2$ for the B and R band, respectively.
Since there is no significant dust extinction for
Mrk 273x (see discussion, \S 4), the inferred low optical luminosity
implies that Mrk 273x is a dwarf galaxy. 

The measured line flux, equivalent width (EW) and full width at half
maximum (FWHM) for all identifiable emission lines are listed
in Table 1. As we can see from Table 1 and Fig. 2,
the high excitation emission lines \OIII\ are
very strong and the ratio of the [OIII] lines to \Hbeta\ is as large as
10. We have also calculated the line ratios,
such as \OIIItwo/\Hbeta, \NII/\Halpha, \SIIthree/\Halpha,
and \OI/\Halpha, and put the results on the diagnostic diagrams
of Veilleux \& Osterbrock (1987).
The positions of these line ratios on the
diagrams leave little doubt that Mrk 273x is a Seyfert galaxy.
The FWHM of the \Halpha\ emission is 15.3\AA,
which corresponds to about 700 \kms. The FWHMs of
other permitted and forbidden lines are similar (cf. Table 1).
These (narrow) widths of the emission lines classify
Mrk 273x as a type 2 Seyfert galaxy
(Osterbrock 1989).

\subsection{The Soft X-ray Properties of Mrk 273x}

We fit the soft X-ray spectrum with single
power-law, thermal bremsstrahlung, black body,
and Raymond-Smith thermal plasma models. Although all these models
are acceptable from the point of view of $\chi^2$, the inferred values
for the absorbing gas column density, $N_{\rm H}$,
for the last 2 models are unreasonably low -- even lower than the
known Galactic contribution. The
best power-law fit has $N_{\rm H}=(4.3\pm2.2)\times 10^{20} {\rm cm^{-2}}$,
and a photon index of $-1.98$,
a typical value for broad-line Seyfert galaxies. Using this index, 
the model dependent soft X-ray flux in the 0.1--2.4 keV
band is
$\fx=1.1 \times 10^{-13} {\rm erg~cm^{-2}~s^{-1}}$.
The corresponding soft X-ray luminosity for Mrk 273x
is $6.3\times 10^{41}$ \ergs, again in the range of
Seyfert galaxies. The intrinsic X-ray luminosity obtained with a
thermal bremsstrahlung fit ($kT \approx 1.6$keV) is smaller by $\approx 5\%$.
For comparison, the B magnitude for Mrk 273 is $B=15.7$ and its
soft X-ray flux is
$\fx=1.25 \times 10^{-13} {\rm erg~cm^{-2}~s^{-1}}$. Therefore,
Mrk 273x is as bright as Mrk 273 in the soft X-ray band,
despite the fact that it is one hundred times less luminous than
Mrk 273 in the optical as well as in the 21cm radio continuum (cf. Fig. 1).

The HRI image (shown as contours in Fig. 1) has a pixel scalelength of
$1.0\arcsec$ and the effective resolution (FWHM) of the Gaussian filtered
image is about $5\arcsec$.
The coordinates of Mrk 273x in the HRI image agree with those
determined from the POSS II plate within the HRI position accuracy.
We have also performed spatial analysis for
the PSPC and HRI data. The resulting radial distributions are
consistent with the companion being a point source.
Since the X-ray observations were carried out in multiple-epochs, we
performed $\chi^2$ tests to see whether the observed counts differs from
a constant average emission for Mrk 273 and its companion.
No significant variability was found for either source during
May and June of 1992.

\subsection{Multi-Wavelength Comparisons of Mrk 273x}

Recently Yun \& Hibbard (1998) obtained a radio continuum map for the
Mrk 273 system using VLA.
Extraordinarily extended radio lobes from Mrk 273 of $\sim 200$ kpc
in length are detected. Mrk 273x has been clearly detected as well with 
a flux $f_\nu=1.45{\rm mJy}$ at (observed frequency) $\nu=1.37$GHz. 
Mrk 273x was unresolved due to the large synthesized beam,
$19\arcsec.6 \times 18\arcsec.4$, but it is consistent with being
a compact point-like radio source. 
Hibbard \& Yun (1998) have also mapped HI in this system. A tidal
HI tail was observed in the southern luminous tail, but not in the much
fainter northeast tail, neither from Mrk 273x. Unless this dwarf
AGN is very HI gas-rich, the non-detection of HI is expected, given 
the small source size and large distance; it is also in
agreement with the small $N_{\rm H}$ value inferred from the X-ray analysis
(see \S 3.2).

These multi-wavelength observations show that Mrk 273x has an
unusual spectral energy distribution. 
The fluxes in the soft X-ray, optical and radio are
$(1.1, 0.15, 2 \times 10^{-4}) \times 10^{-13} {\rm ergs~cm^{-2}~
s^{-1}}$, respectively, where we have estimated
the radio flux using $\nu f_\nu$ at $\nu=1.37$GHz.
The flux ratios are therefore $\fx:\fb:f_{\rm radio}=7:1:1.3\times 10^{-3}$.
Comparing with the sample of Griffiths et al. (1996) from five deep
ROSAT fields show that such peculiar spectral energy distributions
are very rare for AGNs. In particular,
the $\log(\fx/\fb)$ for Mrk 273x is 0.86, which is
only achieved by some BL Lac objects (Stocke et al. 1991), and few AGNs
(Maccacaro et al. 1988). 
For comparison, the corresponding value for Mrk 273,
also a Seyfert 2, is $-$1.18 (its value is probably lowered by the
star formation induced in the merging process). To summarize,
the radio, optical and X-ray observations
all point to the existence of an AGN at the center of Mrk 273x.

\section{Discussion}

We have shown that Mrk 273x is a 
powerful soft X-ray source hosted by a faint dwarf galaxy. In fact, the soft
X-ray luminosity for Mrk 273x is the brightest observed to date for a
dwarf galaxy, more than two orders of magnitude brighter than those
for other nearby dwarf galaxies (see introduction, \S 1).

As discussed in \S 2, Mrk 273x is a faint dwarf galaxy,
with $M_B=-16$ and $M_R=-17.2$. So far, we have neglected
the effect of dust extinction. We now proceed to remedy this.
Following Gebel (1968) and Veilleux \& Osterbrock (1987), 
the extinction in the B band is given by
$
A_B=4\times E(B-V)=3.08\times E_{\beta-\alpha},
$
where $E_{\beta-\alpha} = 2.5 \log[3.10/({\rm H}\alpha/{\rm H}\beta)]$,
appropriate
for AGNs (Miller \& Mathews 1972). Using the \Halpha\ and \Hbeta\ fluxes
listed in Table 1, we find that $A_{B}$ is 
zero for Mrk 273x. The low extinction is in agreement
with the low $N_{\rm H}$ value derived from the X-ray spectral analysis,
and the non-detection of HI in the line emission at 21cm. 
The dwarf nature of Mrk 273x can not only be seen from its faint
luminosity but also from its small size:
its diameter is 7 kpc at most, about a factor of 2 smaller than the upper
limit, 12.8 kpc, for ``true'' dwarf galaxies classified by
Pildis et al. (1997). The HST WFPC2 snapshot
observations show 
a bright nucleus at the center of Mrk 273x; Mrk 273x is therefore
one of the so-called nucleated dwarf galaxies
(Ferguson \& Binggelli 1994). On the other hand,
Mrk 273x is also clearly a Seyfert galaxy. The overall
optical spectrum, in particular the emission
line ratios, of Mrk 273x is typical of Seyfert galaxies. The
soft X-ray properties are characteristic of
Seyfert 2 galaxies. The radio emission is consistent
with a point-source, providing further support for the Seyfert nature of
Mrk 273x. It is, however, unclear
why Mrk 273x is the only X-ray luminous dwarf galaxy detected by ROSAT out of
$\sim 10$ faint objects within 100 kpc of Mrk 273 (cf. Fig. 1). 
A closer examination of the POSS II film (cf. Fig. 1)
and HST images reveals that Mrk 273x is at the tip
of the faint northeast plume from Mrk 273. It is possible that the high
soft X-ray luminosity of Mrk 273x is somehow connected to these features. 

The physical mechanism that produces 
the unusually high soft X-ray luminosity in Mrk 273x is unclear.
Accretion onto a central black hole provides a plausible
explanation (e.g., Lin, Pringle, \& Rees 1988), although some
contribution from starbursts is also possible (e.g., Terlevich et
al. 1992). The X-ray luminosity could also have been boosted by relativistic
effects such as a superluminal jet pointing toward us. Recent numerical
simulations show that major mergers or ``galaxy harassment'' are very
efficient transporting the gas to the centers of interacting galaxies
(Mihos \& Hernquist 1996; Lake, Katz \& Moore 1998).
It is much less clear whether the transported gas forms stars and/or
is accreted onto a (pre-existing) black hole or even forms a black hole.
If the transported gas is
accreted onto a pre-existing black hole, then the mass of the
black hole is $\Mbh \sim 5\times 10^4 M_\odot (0.1/\beta)$, where
$\beta$ is the X-ray luminosity in units of the Eddington luminosity. 
The required accretion rate is
$\dot{M} = \Lx/(\eta c^2)= 10^{-4} (0.1/\eta) M_\odot~{\rm
yr^{-1}}$, where $\eta$ is the conversion efficiency of the accreted mass
into the soft X-ray radiation. The estimates for the mass
and accretion rate are uncertain; the black hole mass is
reminiscent of M32 which has $M_B=-16.6$ and a black
hole mass of $2.3\times 10^6 M_\odot$ (Bender, Kormendy \& Dehnen 1996;
see also Magorrian et al. 1997). 

The mechanism that powers Mrk 273x is unlikely to operate on Mrk 273x alone:
there is likely some population of X-ray luminous 
dwarf AGNs around merging galaxies.
The optical identification of the ROSAT deep survey by Schmidt et
al. (1997) found that out of fifty X-ray sources, four remain
unidentified; they suggest that two of these could have an
unusually large ratio of X-ray to optical flux. In light of the
discovery of Mrk 273x, it may be fruitful to see whether these are also
produced by dwarf galaxies.
In practical terms, Mrk 273x highlights an exciting way to 
detect faint AGNs, and potential black holes at centers of faint galaxies,
by studying the optical counterparts of soft X-ray companions of
Seyfert 2 galaxies and major merging galaxies. If X-ray sources like
Mrk 273x exist in large numbers, they can make a
non-negligible contribution to the X-ray background and the AGN
luminosity function, especially at the faint end.

\acknowledgments
The authors appreciate valuable discussions with Prof.
H.J. Su, Drs. X.W. Cao and H.J. Mo.
Many thanks are due to Drs. M.S. Yun and J.E. 
Hibbard for providing us their recent 21 cm continuum map for the
Mrk 273 field prior to publication. We also thank the helpful comments
by an anonymous referee.
We thank the BATC group for obtaining a deep exposure of the Mrk 231 field.
X.-Y. X. and Z.-G. D. thank MPA for hospitality and the support of
NSFC-DSF exchange program.
YG's research at the Laboratory for Astronomical Imaging in the 
Astronomy Department at the University of Illinois is funded by
NSF grants AST93-20239 and AST96-13999, and by the 
University of Illinois.

\clearpage

\begin{figure}
\plotone{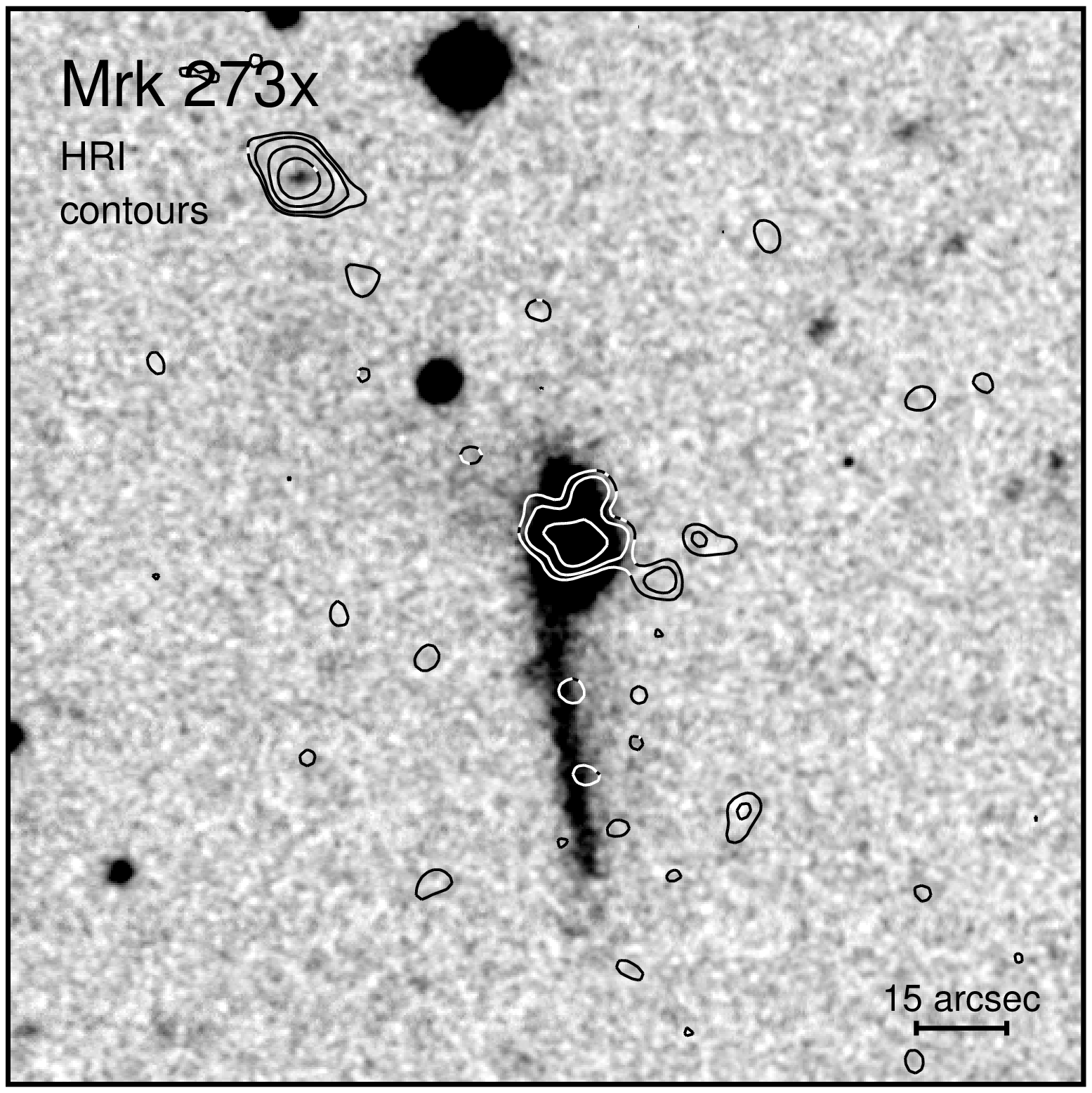}
\figcaption{
Overlay of the soft X-ray contours on the optical image centered on Mrk 273.
The optical image is extracted from an R film POSS II catalog.
The X-ray contours are computed from the ROSAT HRI image and are
$3,5,10,20\sigma$ of the background level,
respectively. There is a bright soft X-ray companion, termed as Mrk 273x,
$1.3\arcmin$ to the northeast of Mrk 273. Clearly
there is a faint dwarf galaxy at the center of the X-ray emission
of Mrk 273x. Approximately ten other faint objects are
identifiable. There is also a (faint) northeast plume 
originating from Mrk 273. Mrk 273x is at the tip of the plume.
}
\end{figure}

\begin{figure}
\plotfiddle{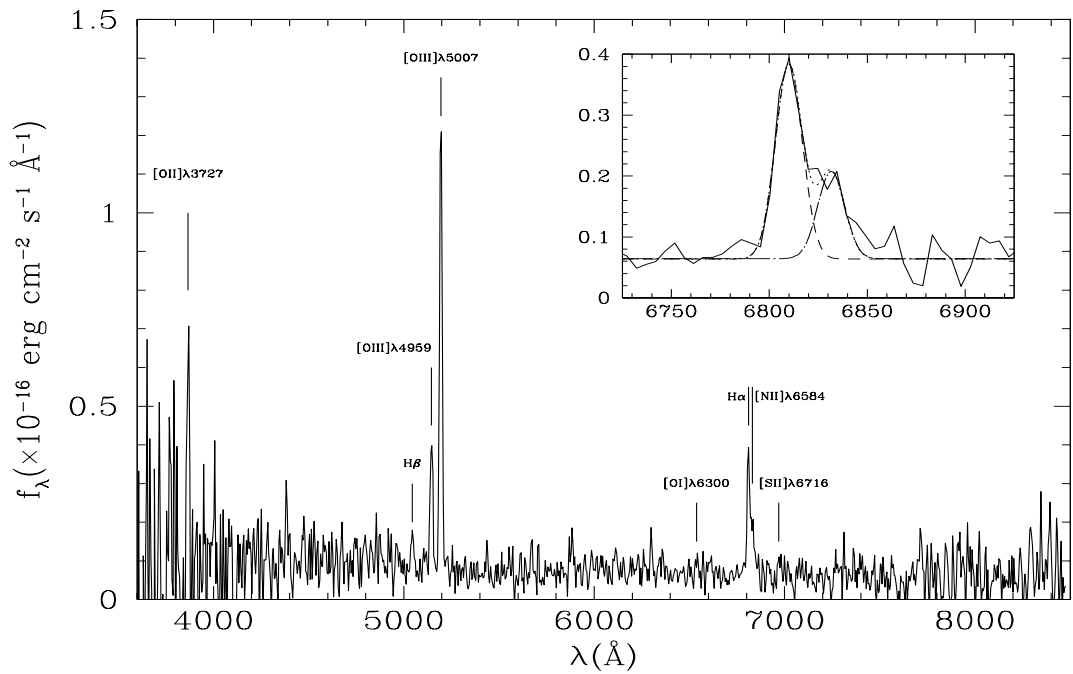}{10cm}{-90}{60}{50}{-250}{288}
\figcaption{
The optical spectrum of Mrk 273x. Some
emission lines are indicated. The strong oxygen lines
are characteristic of Seyfert galaxies. The inset shows a
double-Gaussian fit to the \Halpha\ region. The dashed
(dot-dashed) curve shows the \Halpha\ (\NII) component, the dotted curve
is their sum.
}
\end{figure}

\clearpage

\begin{deluxetable}{llrrll}
\tablecaption{Emission Line Properties for Mrk 273x}
\tablehead{
\colhead{line}	& \colhead{flux} & \colhead{error} & \colhead{EW} &
\colhead{FWHM}}

\startdata
\OII        & 1.05e-15           &  $20\%$    & \nodata   &  15.0  \nl
\OIIIone    & 6.15e-16           &  $10\%$    & 81.8      &  16.5  \nl
\OIIItwo    & 1.94e-15           &  $10\%$    &253.0      &  14.8  \nl
\Hbeta      & 1.90e-16           &  $15\%$    & 27.5      &  18.0  \nl
\OI         & 4.00e-17           &  $100\%$   & \nodata   &\nodata \nl
\Halpha     & 5.60e-16           &  $20\%$    & 87.3      &  16.2  \nl
\NII        & 2.50e-16           &  $20\%$    & 39.1      &  16.2  \nl
\SIIone     & 5.80e-17           &$\geq 30\%$ & \nodata   &\nodata \nl
\SIItwo     & 3.60e-17           &$\geq 30\%$ & \nodata   &\nodata \nl

\tablecomments{
The flux is in units of 
{\rm ergs cm$^{-2}$ s$^{-1}$}
while the EW and FWHM are in units of \AA. The EW of [OII] line
is not given because the spectrum at $\sim 3700$\AA\ is too noisy to
allow a reliable estimate of the continuum.
} 
\enddata
\end{deluxetable}

\end{document}